\documentclass[10pt]{article} 
\usepackage{amsmath,amsfonts,amssymb,amsthm}
\usepackage{mathtools}
\usepackage{graphicx,float}

\usepackage{hyperref}

\usepackage{fullpage}
\usepackage[top=1.9 cm, bottom = 1.9 cm, left = 1.9 cm, right = 1.9cm]{geometry}
\usepackage{cancel}
\usepackage{systeme}

\makeatletter

\newcommand{\Rmnum}[1]{\expandafter\@slowromancap\romannumeral #1@}
\makeatother

\newcommand*{\bpm}{\begin{pmatrix}}
\newcommand*{\epm}{\end{pmatrix}}

\usepackage{bbold} 

\title{A formulation of the relaxation phenomenon for lane changing dynamics in an arbitrary car following model}
\author{Ronan Keane \\  Systems Engineering, Cornell University, Email: (rlk268@cornell.edu) \\ 
H. Oliver Gao \\ Civil and Environmental Engineering, Cornell University,  Email: (hg55@cornell.edu)}
\date{2/26/2021}

\begin{document}
\maketitle 
\section*{Abstract}
Lane changing dynamics are an important part of traffic microsimulation and are vital for modeling weaving sections and merge bottlenecks. However, there is often much more emphasis placed on car following and gap acceptance models, whereas lane changing dynamics such as tactical, cooperation, and relaxation models receive comparatively little attention. This paper develops a general relaxation model which can be applied to an arbitrary parametric or nonparametric microsimulation model. The relaxation model modifies car following dynamics after a lane change, when vehicles can be far from equilibrium. Relaxation prevents car following models from reacting too strongly to the changes in space headway caused by lane changing, leading to more accurate and realistic simulated trajectories. We also show that relaxation is necessary for correctly simulating traffic breakdown with realistic values of capacity drop.


\textit{Keywords:} lane changing, car following, relaxation, traffic microsimulation, LSTM, capacity drop
\pagebreak

\section{Introduction}
\begin{figure}[H] 
\centering 
\includegraphics[ width=\textwidth]{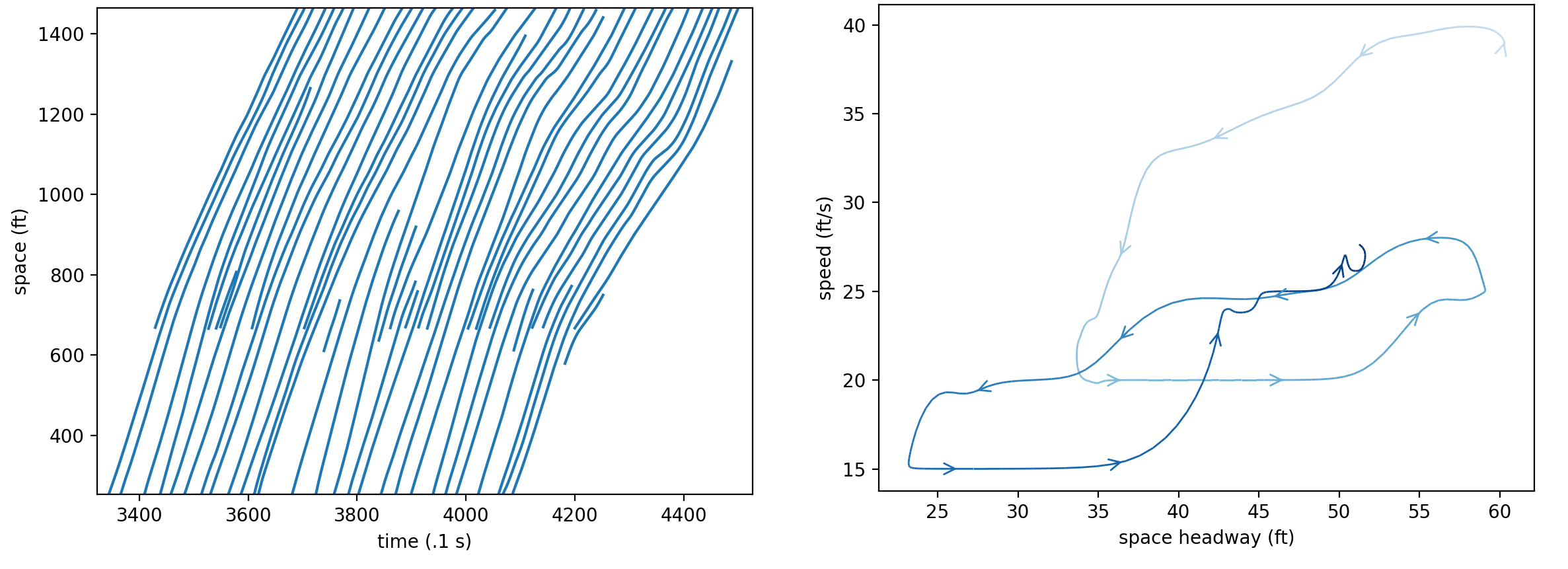} 
\caption{Left panel: space-time plot of vehicle trajectories from lane 6 in the NGSIM I-80 dataset, recorded in congested conditions. There is a merge bottleneck at space = 670 ft. owing to an on-ramp. But despite the abnormally short headways due to merging vehicles, there are no immediate drops in speeds; rather, vehicles gradually relax to a normal spacing (there are also several vehicles which move over to the less congested lanes shortly after merging). \newline
Right panel: speed-headway plot of a typical merging vehicle (id 1336). Darker colors indicate later times, and arrows point in the direction of the trajectory. The vehicle initially merges at a comparatively high speed before transitioning to equilibrium. }  \label{fig1}
\end{figure}

\begin{figure}[H] 
\centering 
\includegraphics[ width=\textwidth]{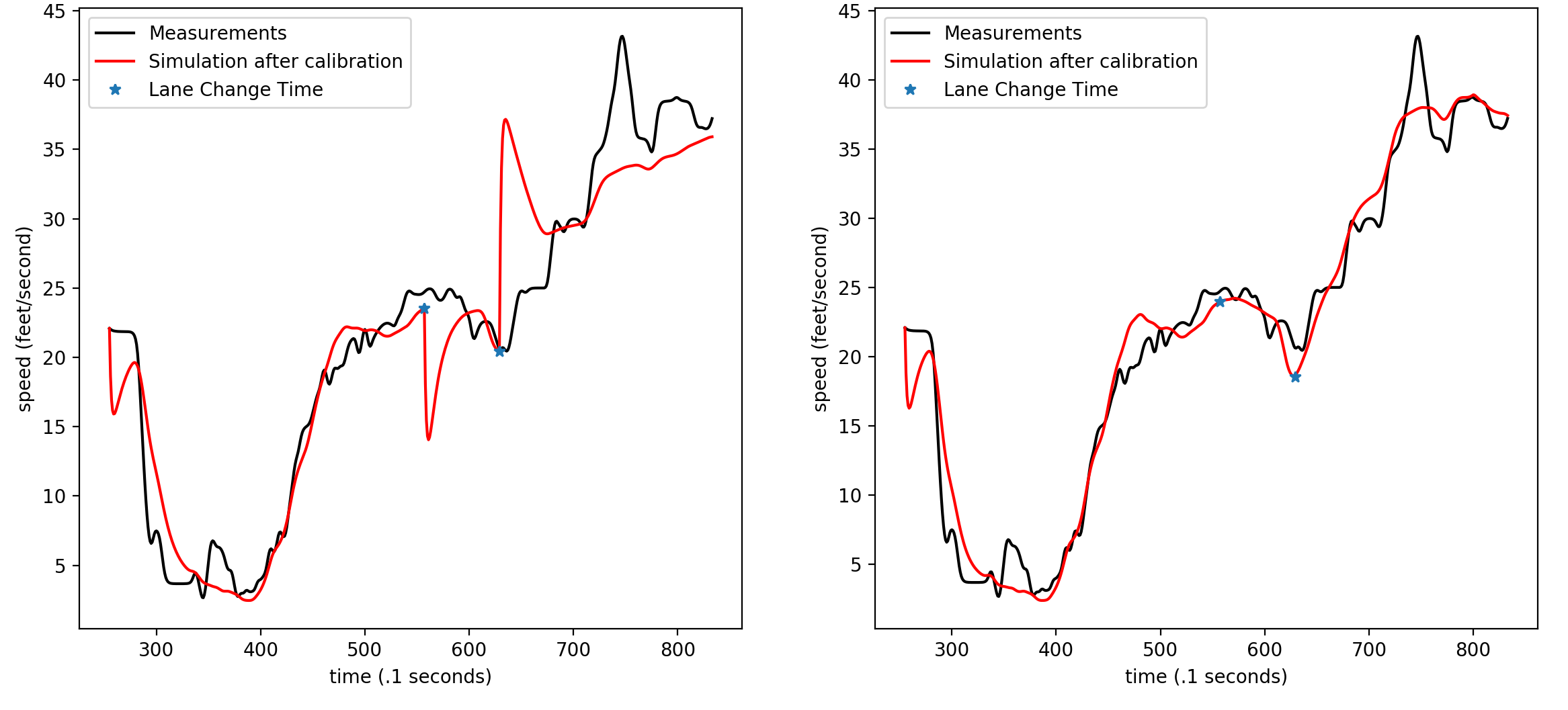} 
\caption{The optimal velocity model (OVM) was calibrated to the trajectory of vehicle 31 in the NGSIM I-80 dataset. Stars indicate the times of the lane changes. The left shows the calibration result when there is no relaxation model--- the model overreacts to the sudden change in headway after a lane change. On the right, the calibration result for OVM with the proposed relaxation model added.} \label{fig2}
\end{figure}

Lane changing (LC) is a vital component of traffic dynamics which plays a role in many observed macroscopic phenomena. Recent works have confirmed that shockwaves on the highway are often initiated by lane changing maneuvers. \cite{105} and \cite{106} both examined the same traffic data from Interstate I-80 in the U.S., with the former concluding that all shockwaves observed in that data originated from disturbances caused by lane changing. The latter concluded that 16 out of 18 traffic oscillations in the data were caused by lane changing. \cite{106} also looked at another study site, and found 12 out of 35 traffic oscillations were caused by lane changing. These results indicate that understanding lane changing dynamics is necessary for understanding congestion. Another example of the importance of lane changing is \cite{40}, where incorporating lane changing dynamics into a kinematic wave model was found to explain the drop in the discharge rate (i.e. capacity drop) frequently observed at the onset of congestion. \\
One well known feature of lane changing dynamics is the so-called relaxation phenomenon. This refers to the observation that drivers are willing to accept abnormally short spacings at the onset of lane changes, and that these short spacings gradually transition (``relax") back to a normal spacing. Car following models react to this short spacing by decelerating too strongly, causing unrealistic behavior. Fig. \ref{fig1} shows some empirical examples of the relaxation phenomenon, and Fig. \ref{fig2} shows an example of the unrealistic behavior which can occur in car following models because of lane changing. \\
This paper proposes a model for the relaxation phenomenon which addresses this unrealistic behavior for an arbitrary model. The relaxation model is meant to augment an existing traffic microsimulation model consisting of both a car following model (which determines the speeds/accelerations of vehicles) and a lane changing model (to determine how/when vehicles should initiate lane changes).
\subsection{Lane changing decisions and lane changing dynamics} \label{sec12}
\begin{figure}[H] 
\centering 
\includegraphics[ width=.9\textwidth]{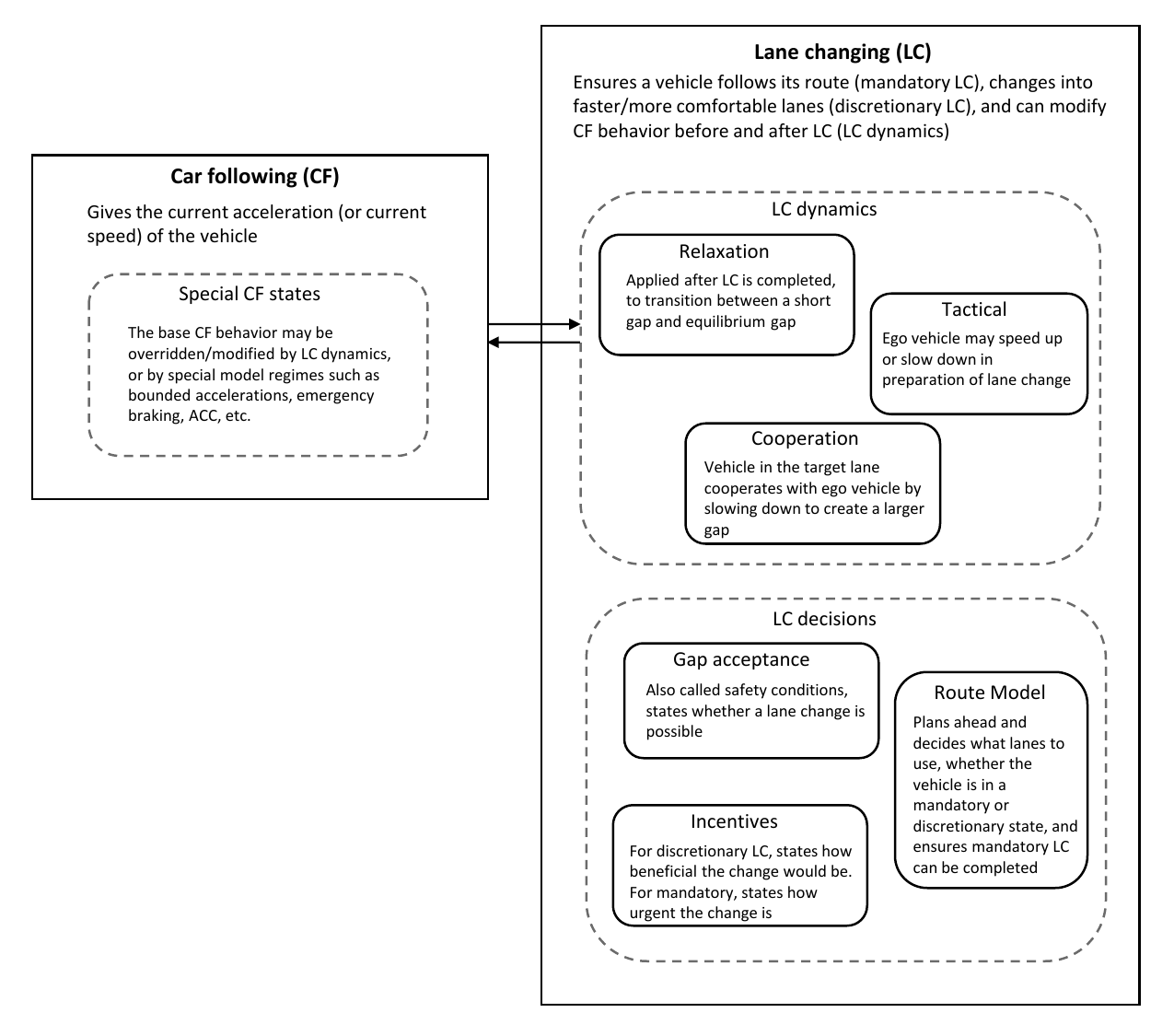}
\caption{High level summary of the different components of a typical lane changing model used in microsimulation.}  \label{cflc}
\end{figure}

A complete lane changing (LC) model consists of several interacting parts (see Fig. \ref{cflc}). One important distinction is between lane changing decisions (the decision processes behind lane changing, which control whether a vehicle will change lanes), and lane changing dynamics (changes to the driving dynamics applied as a result of lane changing). A typical lane changing model (for example, \cite{34}, \cite{1}, \cite{sumolc}, \cite{toledolc}, \cite{gippslc}, \cite{hidaslc}) can be summarized as consisting of the following parts
\begin{enumerate}
\item A gap acceptance model or `safety condition', which determines whether there is sufficient space in the target lane to initiate the LC. The gap required usually depends on both the speeds and positions of the leading and following vehicles in the target lane (the preceding and proceeding vehicles if one was to change lanes). 
\item Rules for both discretionary and mandatory LC. In a discretionary LC, the ego vehicle changes lanes to obtain a more favorable traffic condition. Typically there is some `incentive' which promotes changing into faster lanes; passing rules should also be taken into consideration. In  mandatory LC, the ego vehicle must change lanes to stay on their desired route. This includes planning to decide which lanes the vehicle should use, and rules for handling LC which need to be completed urgently.
\item Lane changing dynamics, which modify the vehicle's CF model in various ways. The most common example is cooperation, where a vehicle in the target lane slows down with the intention of creating a gap for the ego vehicle to use. For an urgent mandatory LC, cooperation may be forced in the target vehicle, simulating aggression of the ego vehicle. There may also be rules (sometimes referred to as a tactical model) for the LC vehicle to either accelerate or decelerate in order to find a suitable gap in the target lane to use. Whereas a tactical/cooperation model is applied before an LC, a relaxation model is applied after the LC, when vehicles must adjust to their new driving situations.
\end{enumerate}
Much of the existing literature concentrates on LC decisions, particularly for discretionary LC. But with a discretionary LC model, it is only possible to model a straight highway section. For an arbitrary highway network, consisting of on/off ramps, bottlenecks, and weaving sections, both mandatory LC and LC dynamics need to be considered. \\
Examining the literature describing complete LC models, we found that although LC dynamics typically include cooperation, a formulation of relaxation is often lacking. Relaxation is critical for modeling short gaps. With relaxation, short gaps can be accepted, because vehicles will adjust to equilibrium gradually without overreacting; without a relaxation model, vehicles would have to wait until a larger gap occurs naturally or is created by the cooperation/tactical model. In \cite{mergingbehavior}, it was found that existing lane changing models were not able to describe many of the LC observed in real data because the models lacked an implementation of the relaxation phenomenon. \\
In a recent LC model literature review, \cite{107} identified one literature gap as a general model of the relaxation phenomenon which can be combined with an arbitrary car following/microscopic model. For existing relaxation models, they are either formulated for macrosimulation or are formulated for specific microsimulation models. \cite{31}, \cite{41}, and \cite{108} all develop formulations of the relaxation phenomenon, but these works are only interpretable in the setting of kinematic wave theory/the LWR model. \cite{82} developed a relaxation model which only applies to the Newell car following model. \cite{81}, \cite{80}, \cite{1}, \cite{32} also developed formulations of relaxation for car following models. Those works incorporated relaxation by changing some parameter in the model in order to temporarily accept the short spacing. The present work differs because instead of changing model parameters, the input to the model (i.e. the headways, speeds) is altered. This creates a formulation of the relaxation phenomenon which can be easily applied to an arbitrary car following model. 

\section{The Relaxation Model} \label{sec2}
\begin{figure}[H] 
\centering 
\includegraphics[ width=.5\textwidth]{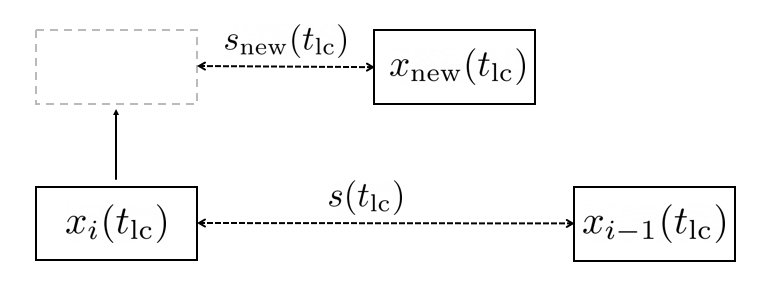} 
\caption{At time $t_{\rm lc}$, vehicle $x_i$ initiates a lane change and begins following $x_{\rm new}$ instead of $x_{i-1}(t_{\rm lc})$.}  \label{fig3}
\end{figure}

\begin{figure}[H] 
\centering 
\includegraphics[ width=\textwidth]{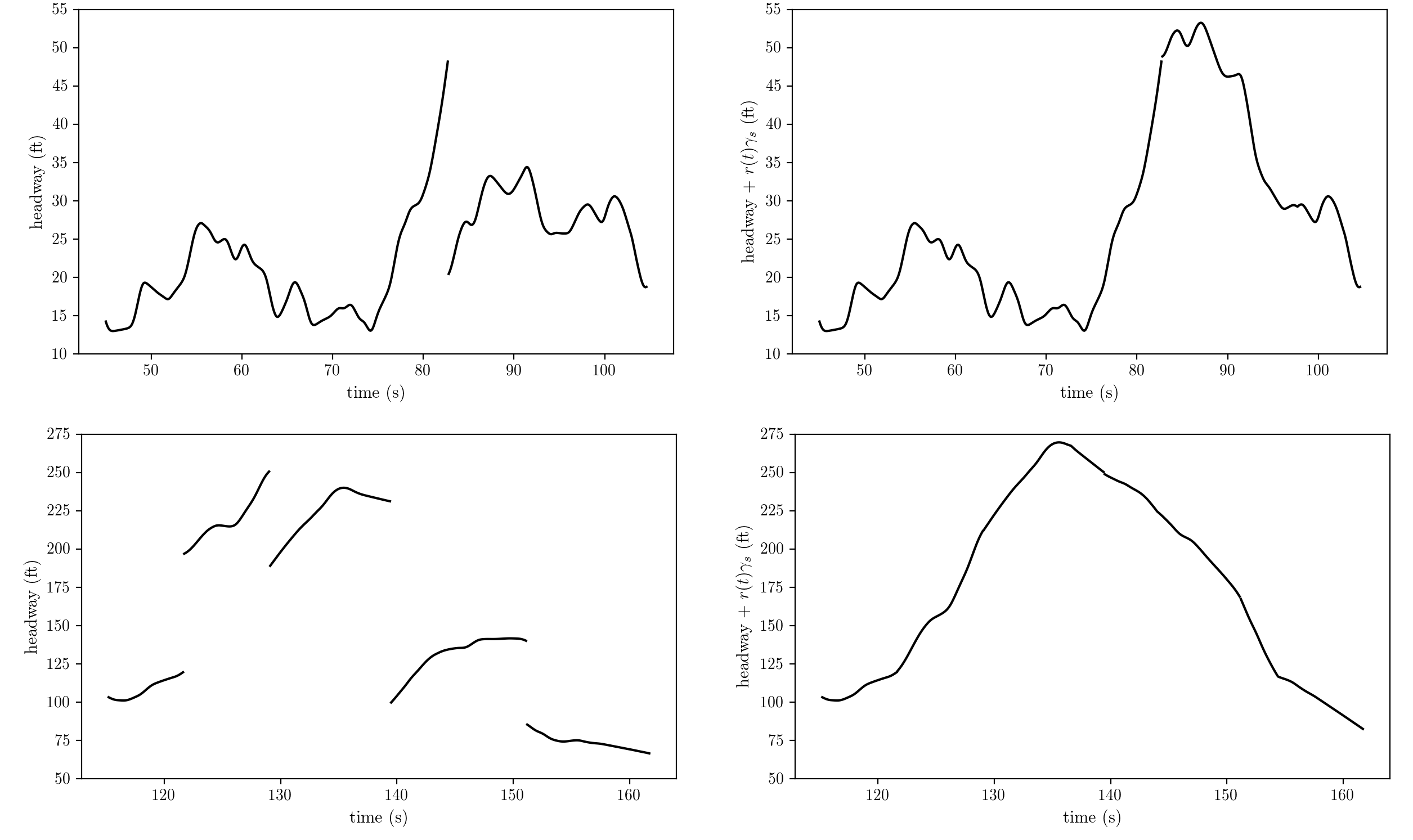} 
\caption{Top left shows the recorded headway for vehicle 93, and bottom left shows the recorded headway for vehicle 320. The gaps in headway are caused by lane changing maneuvers. The right shows the relaxed headway for their respective vehicles, using a value of $c = 15$ seconds. The relaxed headway is used in place of the normal headway as an input to the car following model. } \label{fig4}
\end{figure}

Consider the second order ODE car following model $h$ 
\begin{align*} 
\ddot x_i(t) = h(s(t), \ \dot x_{i-1}(t), \ \dot x_i(t)) \tag{1}\label{1}
\end{align*}
where $s(t)$ is the (space) headway at time $t$, $\dot x_{i-1}(t)$ is the speed of the lead vehicle at time $t$, and $\dot x_{i}(t)$ is the speed of following vehicle at time $t$. The model with relaxation added is 
\begin{align*} 
\ddot x_i(t) = h(s(t) + r(t)\gamma_s, \ \dot x_{i-1}(t) + r(t)\gamma_v, \ \dot x_i(t)) \tag{2} \label{2}
\end{align*}
where $r(t)$ is the relaxation and $\gamma_s, \gamma_v$ are the relaxation amounts for headway and speed.
\begin{align*} 
& r(t) = \begin{cases} 
1 - \dfrac{1}{c}(t - t_{\rm lc}) & t_{\rm lc} < t < t_{\rm lc} + c \\ 
  0 & \text{otherwise} \\ 
  \end{cases} \\
& \gamma_s =  s(t_{\rm lc}) - s_{\rm new}(t_{\rm lc}) \\
& \gamma_v =  \dot x_{i-1}(t_{\rm lc})  - \dot x_{\rm new}(t_{\rm lc}) \tag{3}\label{3}
\end{align*}
Where $t_{\rm lc}$ is the time of the lane change, $s_{\rm new}$ is the headway with respect to the new leader, and $\dot x_{\rm new}$ is the speed of the new leader. This formulation introduces one parameter, $c$, which defines the relaxation time. See Fig. \ref{fig3} for a visual explanation. Note we treat $t_{\rm lc}$ as the last timestep the vehicle follows the old leader.
As opposed to existing relaxation models which temporarily modify model parameters, our relaxation model is based on modifying the headway/lead vehicle speed. This formulation avoids the sudden jumps in those quantities which occur when $x_{i-1}$ changes due to a lane change. See fig. \ref{fig4} for some examples of the jumps in unmodified headway and what the relaxed headway looks like. There are two main benefits to our formulation. First, because we change the inputs and not a model parameter, the relaxation can be applied to an arbitrary model, including different classes of car following models and also nonparametric models. Second, the model $h$ will be continuous even during lane changes. This helps to ensure smooth transitions without overreactions to lane changes.
\subsection{Mathematical analysis of the relaxation model}\label{math}
\begin{figure}[H] 
\centering 
\includegraphics[ width=\textwidth]{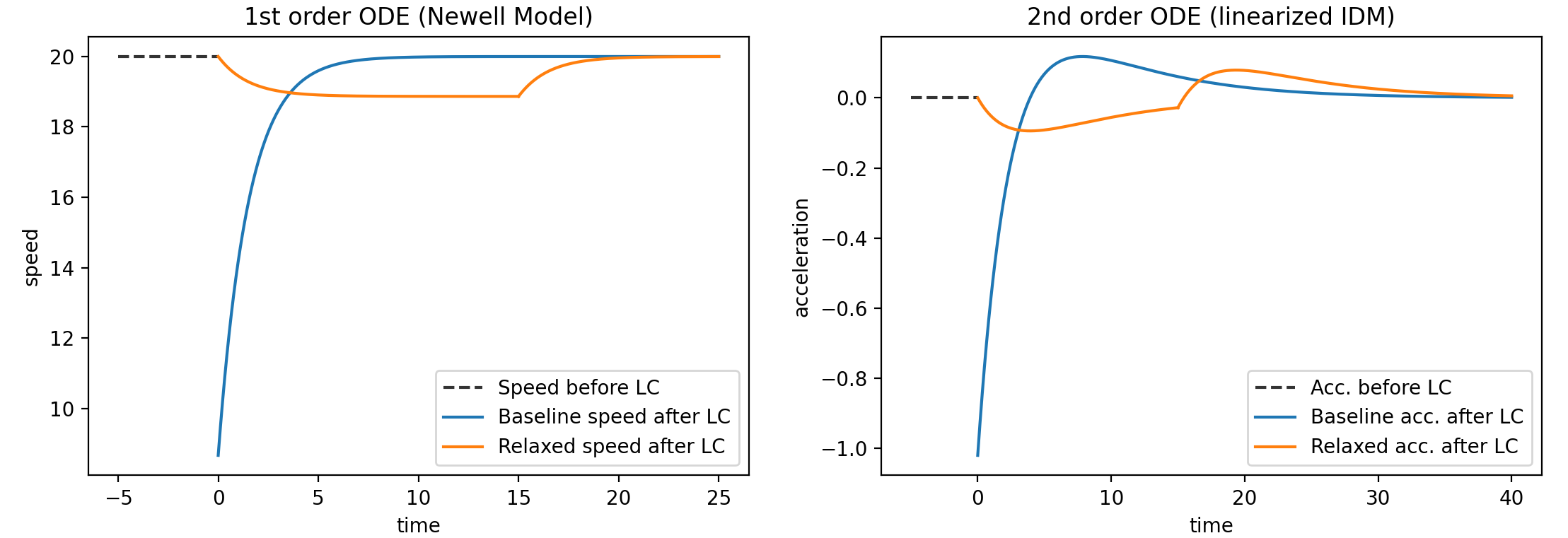} 
\caption{The following vehicle was initially at equilibrium and changed leaders at $t=0$, resulting in a relaxation amount of $\gamma_s = 17$ ($\gamma_v = 0$). For the baseline model with no relaxation, there are strong over-reactions to the lane change.}  \label{fig5}
\end{figure}

We can obtain closed form solutions for linear models in order to analyze the effect of relaxation for a vehicle adjusting to a non equilibrium spacing. Consider the following first order and second order linear car following models 
\begin{align*} 
& \dot x_i(t) = \beta_1 ( s(t) - \beta_2 ) \label{4}\tag{4} \\
& \ddot x_i(t) = \beta_1 s(t) + \beta_2 \dot x_i(t) + \beta_3 \dot x_{i-1}(t) + \beta_4 \label{5}\tag{5}
\end{align*}
where the $\beta$ are model parameters. The Eq. \eqref{4} is equivalent to the Newell model, where $\beta_2$ is the jam spacing and $\beta_1\beta_2$ is the wave speed. We used the parameters $\beta_1 = 2/3$ and $\beta_2 = 2$ (with units of 1/seconds and meters respectively). Eq. \eqref{5} is a generic linear second order car following model, and we obtained the parameters $\beta_1 = .06$, $\beta_2 = -.55$, $\beta_3 = .45$, $\beta_4 = .14$ by linearizing the IDM. \\
We solve for the trajectory of both models experiencing a lane change at $t_{\rm lc} = 0$ where the previous leader $x_{i-1}$  and new leader $x_{\rm new}$ both have constant speeds of 20 m/s. The initial position of the vehicles was chosen so that $\gamma_s = 17$ m for both models, and the follower was assumed to be in equilibrium prior to the change, so that $\dot x_i(t) = 20$ before the change. Fig. \ref{fig5} shows the resulting speed profiles (for Newell) and acceleration profiles (for linearized IDM) using a relaxation time of $c = 15$.
Turning our attention to the Newell model, the speed profile for the baseline is 
\begin{align*} 
\dot x_i(t) = v - \gamma_s \beta_1 \exp( -\beta_1 t)
\end{align*}
where $v$ is the (constant) speed of the previous/new leader. This means that the follower is initially $\gamma_s \beta_1$ m/s away from the equilibrium speed of $v$, and that this difference decreases at an exponential rate of $\beta_1$. Note that for the baseline model, the time needed to reach equilibrium is proportional to $1/\beta_1$. The speed profile for the relaxed model is 
\begin{align*} 
\dot x_i(t) = \begin{cases} 
  -\dfrac{\gamma_s-cv}{c} + \dfrac{\gamma_s}{c}\exp(-\beta_1 t)  &  0 <= t < c \\ 
  v - \alpha_1 \exp(-\beta_1 t) & c <= t \\ 
  \end{cases} 
\end{align*}
where $\alpha_1$ is a constant. Assuming that the relaxation time $c$ is significantly longer than $1/\beta_1$, then the speed will adjust to $-(\gamma_s - cv)/c$ and hold that constant speed for the duration of the relaxation. When the relaxation ends, the model will behave like the baseline model, but the initial speed difference will be $\gamma_s/c$ instead of $\gamma_s \beta_1$. \\
Define the time to equilibrium (TTE) as the length of time needed to adjust within $\delta$ of the equilibrium speed.
\begin{align*} 
\rm TTE = \dfrac{1}{\beta_1} \log \left( \dfrac{\beta_1 \gamma_s}{\delta} \right)  \quad \text{ for baseline }\\
\rm TTE = c + \dfrac{1}{\beta_1} \log \left( \dfrac{\gamma_s}{\delta c} \right) \quad \text{ for relaxation}
\end{align*}
Note that the TTE is not the same as the relaxation time parameter $c$--- it also depends on the car following parameter $\beta_1$ (or more generally, for a higher order car following model the TTE will be inversely proportional to the largest root of the characteristic equation). The relaxed TTE is also not exactly equal to the baseline TTE $+ \ c$: it is shorter because when the relaxation ends the vehicle is already closer to equilibrium. \\
We also define the deceleration time (DT) as the length of time that the vehicle spends decelerating after the LC. For the Newell model, DT is simply equal to $c$ and the unrelaxed model has DT$=0$. For second order models, the unrelaxed model has some DT $> 0$, and the relaxed DT is approximately $c$ greater than the unrelaxed DT. \\
Overall this analysis shows that for a leader with constant speed, the relaxed Newell follower will essentially adjust to the new equilibrium at a constant speed. It's not true that the model will reach equilibrium after $c$ time; to characterize a vehicle's adjustment towards equilibrium, we suggest calculating/estimating the TTE and DT values.
One important finding is that different models have different baseline TTEs/DTs, meaning it is not appropriate to give different models the same $c$ values. For example, the Newell model with realistic parameters has a TTE $\approx$2--4 seconds and DT$=0$, whereas the IDM has a TTE $\approx$10--20 seconds and DT$\approx$1-2 seconds. The $c$ value should ideally be calibrated with the rest of the parameters. Alternatively, if one has observed (i.e. from traffic data) approximate values for TTE or DT, it is possible to calculate $c$ analytically.

%
\subsection{Extra details}
Calculating the relaxation amounts $\gamma_s, \gamma_v$ in Eq. \eqref{3} requires both the old leader $x_{i-1}(  t_{\rm lc})$ and new leader $x_{\rm new}$. But for merging vehicles, there is typically no old leader as vehicles are on the end of the on-ramp/lane. In that case, we calculate the relaxation amounts as 
\begin{align*} 
& \gamma_s = s_{\rm eql}(\dot x_i(t_{\rm lc}) ) - s_{new}(t_{\rm lc}) \\
& \gamma_v = \dot x_i(t_{\rm lc}) - \dot x_{\rm new}(t_{\rm lc})
\end{align*}
where $s_{\rm eql}(v)$ is the equilibrium headway of vehicle $i$ corresponding to speed $v$. \\
Some vehicles may change lanes again before their current relaxation ends. In these cases, the relaxations are added together (and each lane change has its own $\gamma, r(t)$ associated with it). \\
Since positive values of $\gamma_s$ increase the headway, it is possible that the relaxation model can cause collisions in a model that is otherwise accident free. In our experiments, we found that this occurred rarely, only in cases where the new leader suddenly slows down shortly after the change (e.g. due to a shockwave that arrives in the new lane). To prevent this we safeguarded the relaxation in the following way
\begin{align*} 
\rm{z} = \dfrac{\max(s(t) - s_j - \alpha \dot x_i(t), \epsilon)}{\dot x_i(t) - \dot x_{i-1}(t) } \\
r(t) = \begin{cases} 
 r(t)\left( \frac{z}{\beta} \right) &  0 < \rm{z} < \beta \\ 
 r(t) & \rm otherwise \\ 
  \end{cases}
\end{align*}
Where $s_j$ is the jam spacing of the CF model, $\epsilon$ is some small positive constant, and $\alpha, \beta$ are two new safeguard parameters. We took $\alpha = .6$ and $\beta = 1.5$ seconds and did not calibrate those values.  \\
Lastly, we apply the relaxation to all vehicles which experience a leader change. This means that a single lane change results in relaxation being applied to 3 vehicles. There is the lane changing vehicle $x_i$, the old follower (the vehicle following $x_i$ at $t_{\rm lc}$) and the new follower (vehicle following $x_{\rm new}$  at $t_{\rm lc}$). This is slightly different from existing models such as \cite{80} or \cite{41} which do not apply relaxation to the old follower. Also, we apply the relaxation whether the $\gamma_s$ is positive or negative. In those previous papers, they only consider relaxation for $\gamma_s$ being positive (i.e. headway decreases after lane change). 

\section{Microscopic Validation}
\subsection{Models}
To test that the relaxation model can be used with an arbitrary parametric model, we applied it to the optimal velocity model (OVM) \cite{52}, intelligent driver model (IDM) \cite{12}, and Newell model \cite{13}. As a benchmark for existing relaxation models, we considered \cite{41} which formulates a 1 parameter relaxation model for Newell, and \cite{80} which formulates a 2 parameter relaxation model for IDM. \\
The parametrization for OVM is 
\begin{align*} 
& \ddot x_i(t) = c_4(V(s(t)) - \dot x_i(t)) \\
 & V(s) = c_1[ \tanh(c_2 s - c_3 - c_5) - \tanh(-c_3)]
\end{align*}
where $V(s)$ is the optimal velocity function with shape parameters $c_1, c_2, c_3, c_5$. The maximum speed is $c_1(1-\tanh(-c_3))$, $c_5/c_2$ is the jam spacing, and $c_4$ controls the strength of acceleration. \\
The parametrization for IDM is 
\begin{align*} 
\ddot x_i(t) = c_4\left( 1 - \left( \dfrac{\dot x_i(t)}{c_1} \right)^4  - \left( \dfrac{s^*}{s(t)} \right)^2\right) \tag{6}\label{6} \\
s^*  = c_3+c_2\dot x_i(t) + \dfrac{\dot x_i(t) (\dot x_i(t) - \dot x_{i-1}(t))}{2 \sqrt{c_4 c_5}}  
\end{align*}
where $s^*$ is the desired headway. $c_1$ is the maximum speed, $c_2$ is the desired time headway, $c_3$ is the jam spacing, $c_4$ is the acceleration and $c_5$ is the comfortable deceleration. \\
In \cite{80} their relaxation model for IDM works by modifying the time headway parameter $c_2$. Let $T^*(t)$ be the relaxed time headway. In that paper, the initial value for time headway is set to $\min(c_2, \ d \cdot T_{\rm min} + (1-d)c_2 )$ where $d$ is the `desire' which changes based on the current lane changing model incentives, and $T_{\rm min}$ is the minimum possible time headway. The time headway is then updated as
\begin{align*} 
T^*(t + \Delta t) = T^*(t) + (c_2 - T^*(t)) \dfrac{\Delta t}{\tau}
\end{align*}
where $\tau$ is the relaxation time parameter, and $\Delta t$ is the simulation time step. Thus their formulation had 2 relaxation parameters ($T_{\rm min}, \tau$), and is also integrated with the parameters of the lane changing model through the desire. Since we only want to look at car following dynamics in this section, we have set the initial value for time headway to $\min(c_2, T_0)$ where $T_0$ is the new relaxation parameter. This retains a similar degree of freedom while removing the need for jointly calibrating the specific lane changing model used in that paper. \\
The parametrization for Newell is 
\begin{align*} 
x_i(t + \tau) = \min(x_i(t) + v_f \tau, x_{i-1}(t) -  l_{i-1} - \delta)
\end{align*}
where $\delta$ and $\tau$ are the space and time shift parameters, and $v_f$ is the free flow speed (also a parameter). $l_{i-1}$ is the length of the lead vehicle. \\
In \cite{41} their relaxation model for Newell is formulated as 
\begin{align*} 
& x_i(t) = \min( x_i(t - \Delta t) + \min(v_f, \dot x_i(t - \Delta t) + a \Delta t ) \Delta t, \ x_i(t - \Delta t) + \dot x_{i-1}(t)\Delta t - \dfrac{\Delta N(t)}{K(\dot x_{i-1}(t))} ) \\
& K(v) = \dfrac{\omega \kappa}{v + \omega}
\end{align*}
where $\omega, \kappa, v_f$ are the parameters, and the relaxation is controlled by $\Delta N(t)$ and its single parameter $\epsilon$ (refer to the paper for the equations used to initialize and update $\Delta N$). Note that this formulation of Newell is equivalent to the previous one when $\Delta N$ is 1, $\omega = \delta/\tau$, $\kappa = 1/\delta$, and $\tau = \Delta t$. The only exception is there is also a maximum acceleration $a$, which was set to 3.4 $m/s^2$ and not calibrated.
\subsection{Methodology}
In this section we only consider calibrating the car following and relaxation model parameters to vehicle trajectory data. There is no lane changing model, so the lane changes and leader/follower relationships are taken from the data. This allows the simulated trajectories to be directly compared to the measured trajectories at an individual vehicle level. Each vehicle has its own individual parameters which represent only that particular vehicle. \\
For any given vehicle to be calibrated, the lead vehicle trajectories and initial condition of the vehicle are taken from the data. Then the simulated vehicle trajectory is generated using given the car following/relaxation parameters, and the mean squared error (MSE) in position between the simulation and measurements is calculated. A genetic algorithm \cite{73} then finds the parameters which minimize the MSE. \\
We used the reconstructed NGSim data \cite{29} as the source of data, which contains 2037 vehicle trajectories total. Of those, 4 vehicles do not have their lead vehicles in the dataset and cannot be calibrated. Of the remaining 2033, 1101 vehicles change lanes themselves, or have a leader which changes lanes. There are also 158 vehicles which merge from the on-ramp onto the highway.
\subsection{Results} \label{parametric results}
\begin{figure}[H] 
\centering 
\includegraphics[ width=\textwidth]{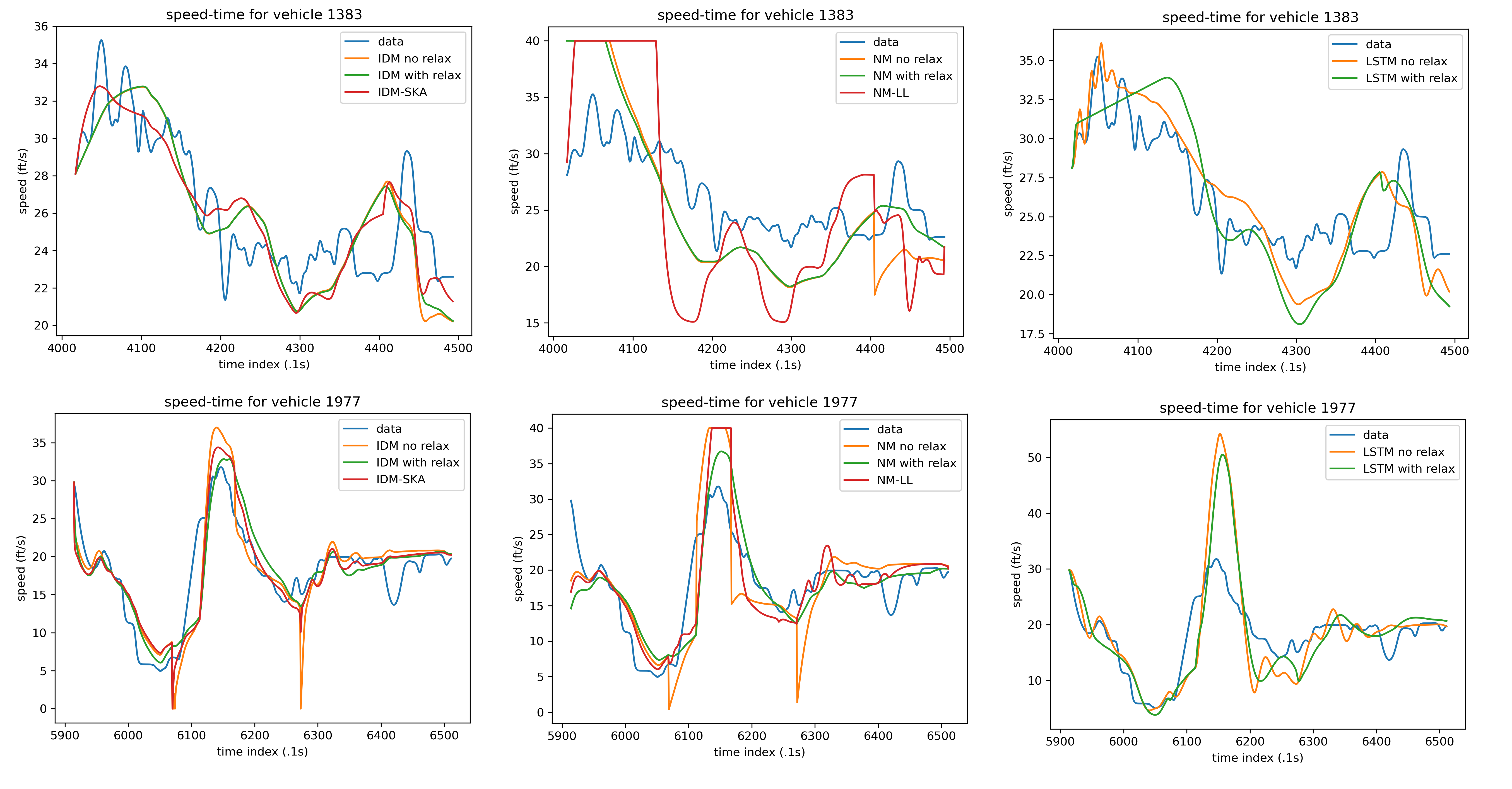} 
\caption{Speed time plots of calibrated trajectories (NM = Newell Model). Many of the Newell trajectories (e.g. middle top/bottom panels) appear unrealistic because of strong accelerations after lane changes. The bottom left panel shows an example of this same effect for the IDM no relax and IDM SKA.} \label{speedfig}
\end{figure}

`Relax' refers to the calibrated results for the car following model with relaxation applied to all the lane changing and merging vehicles. `No relax' are the calibrated results for all the lane changing and merging vehicles, but with no relaxation added. `No LC' are the results for only the vehicles with no lane changing. IDM SKA is the relaxation model due to \cite{80} and Newell LL is the relaxation model due to \cite{41}. Since SKA uses two relaxation parameters, we also tested using our relaxation model with 2 parameters, which is labeled as `IDM relax 2p'. In that formulation, there are two separate relaxation times for the headway/speed relaxation amounts $\gamma_s, \gamma_v$. \\
For the metrics, MSE is the mean squared error averaged over all vehicles, `MSE many LC' is the MSE averaged only over vehicles with 3 or more lane changes, and `MSE merges' is the MSE averaged over merging vehicles. `MSE near LC' is the MSE of trajectories in the 10 seconds after a lane change, averaged over all lane changing events. For all those metrics, we report the mean/median/standard deviation of the distributions of MSE.  \\
Realistic acc. (acceleration) is a measure of whether the calibrated trajectory has any large accelerations/decelerations which were not recorded in the data. This is meant to be a measure of whether the model tends to overreact to lane changes. For each vehicle, we calculate the maximum allowed acceleration/deceleration as $\max ( 4 \rm{m/s}^2, 1.1\max( \hat{\ddot{x}}_i)\ )$, $\min ( -6 \rm{m/s}^2, 1.1\min(\hat{\ddot{x}}_i) \ )$ respectively, where $\max( \hat{\ddot{x_i }})$ is the empirical observed maximum acceleration. For virtually all vehicles, this results in the acceleration bounds of $[4, -6]$ which confirms the plausibility of those bounds. Then the calibrated trajectory is either classified as ``realistic'' or ``unrealistic'' if the simulation has any timesteps with accelerations which are outside those bounds.
\begin{table}[H]
\begin{tabular}{|l|l|l|l|l|l|l|}
\hline
                    & MSE ($m^2$) & MSE near LC & Realistic Acc.  & MSE many LC & MSE merges \\ \hline
IDM relax   & 3.33/1.93/7.0   & 5.34/1.86/14.4  & 77\% & 5.84/2.93/14.3 & 1.39/.76/2.0 \\ \hline
IDM no relax    & 4.41/2.50/9.6   & 7.45/2.66/24.2  & 57\% & 8.20/3.45/20.0 & 2.00/.99/3.5  \\ \hline
IDM no LC    & 1.87/1.22/2.0   & -  & 94\% & - & -  \\ \hline
IDM SKA    & 3.07/1.71/6.6   & 4.50/1.55/13.4  & 62\% & 5.30/2.43/13.5 & 1.16/.52/2.0  \\ \hline
IDM relax 2p   & 2.89/1.61/6.7  & 4.23/1.49/12.3  & 77\% & 4.90/2.44/13.7 & .92/.56/1.0  \\ \hline
OVM relax & 4.68/2.83/8.6   & 7.10/2.88/18.5  & 58\% & 7.82/4.64/16.3 & 1.73/.91/3.2  \\ \hline
OVM no relax    & 7.98/4.90/14.2   & 13.98/6.34/32.8  & 39\% & 14.73/7.32/29.1 & 4.08/2.23/5.3  \\ \hline
OVM no LC    & 2.90/1.80/3.7   & -  & 74\% & - & -  \\ \hline
Newell relax   & 8.47/4.57/17.2   & 11.37/4.44/29.3  & 64\% & 13.17/6.38/20.6 & 3.11/1.77/4.4\\ \hline
Newell no relax    & 16.03/9.09/26.2   & 27.74/13.75/50.7  & 7\% & 25.90/14.56/37.4 & 10.32/5.49/14.1  \\ \hline
Newell no LC   & 5.73/3.35/8.7   & -  & 84\% & - & -  \\ \hline
Newell LL  & 11.39/6.27/21.3   & 18.68/7.07/58.5  & 39\% & 18.80/9.54/38.6 & 5.66/2.99/9.3  \\ \hline
\end{tabular}
\caption{Calibration results of the unrelaxed/relaxed parametric models.} \label{table 1} 
\end{table}
Out of the models tested, IDM consistently performed the best over all metrics, which is a result other papers have also had when comparing different car following models. By contrast, Newell performed the worst, except for the realistic acceleration metric, for which OVM performed the worst. However it is also true that Newell has 3 parameters (+1 with relaxation) compared to IDM/OVM which have 5 (+1 with relaxation, or +2 for SKA and relax 2p). \\
From the mean/median/standard deviation, it is clear that all the distributions are skewed right, with certain vehicles which have high MSEs. These vehicles which contribute the high MSEs tend to be the same across all the different models. Often, these are vehicles which have many lane changing events over their trajectory--- consider that the `MSE many LC' is significantly higher than the MSE for all models. In general, the MSE for lane changing vehicles is significantly higher than vehicles with no lane changing, even after adding relaxation. The MSE near LC metric confirms that this is due to the lane changing itself, since that metric is significantly higher than the average MSE. \\
As for merging vehicles, we found their average MSE were significantly lower than the average lane changing vehicle. This is because the merging vehicles in the dataset tended to only experience one lane change (the merge itself), and then stayed in the most congested rightmost lane. Thus, merging vehicles tended to have few lane changes, and stay at relatively constant speeds compared to the average lane changing vehicle, resulting in their lower MSE. \\
For all models tested, the `realistic acceleration' for lane changing vehicles is lower than for non lane changing vehicles. Note though that even though the metric is called ``realistic'', a trajectory can still have accelerations which fall outside the [4, -6] m/s$^2$ bounds and still be realistic. In general we found that nearly all of the trajectories for IDM using either SKA or the relaxation model appeared qualitatively realistic, but there they have some short time periods with acceleration just outside the [4, -6] bounds and therefore contribute against the metric. On the flip side, there were many trajectories for OVM and Newell that still had severe and clearly unphysical acceleration spikes, even after adding relaxation. \\
Comparing Newell relax to the existing relaxation model Newell LL, our formulation had significantly better results in every category. IDM relax (+1 parameter) did slightly worse than IDM SKA (+2 parameters) in every category except realistic acceleration, where our model did significantly better. But when we also used 2 relaxation parameters, our model is again outperforming the existing model in every category. As for average relaxation values, we found an average/median of 21.1/8.7 for IDM, and 26.5/15.3 for Newell. The larger relaxation parameters for Newell are to be expected and confirm the analysis of section \ref{math}. \\ 
Some randomly chosen examples of speed time series are shown in Fig. \ref{speedfig}.
Overall, we summarize the results for this section as follows
\begin{itemize}
\item The MSE for lane changing vehicles is significantly higher than for non lane changing vehicles, and base car following models with no relaxation tend to overreact to lane changes. 
\item Our relaxation model is comparable or better than existing model specific relaxation models, in terms of level of fit and ability to prevent overreaction.
\item Even after adding relaxation, MSE for lane changing vehicles is still significantly higher than non lane changing vehicles. This points to a current literature gap which is how to properly model and calibrate lane changing at a microscopic level. Future research should consider models which use leaders/followers in neighboring lanes, and develop methodology for calibrating not only relaxation but also tactical/cooperative lane changing models at a microscopic level.
\end{itemize}
\subsection{Validation with a neural network model}
\begin{figure}[H]
\centering 
\includegraphics[ width=.75\textwidth]{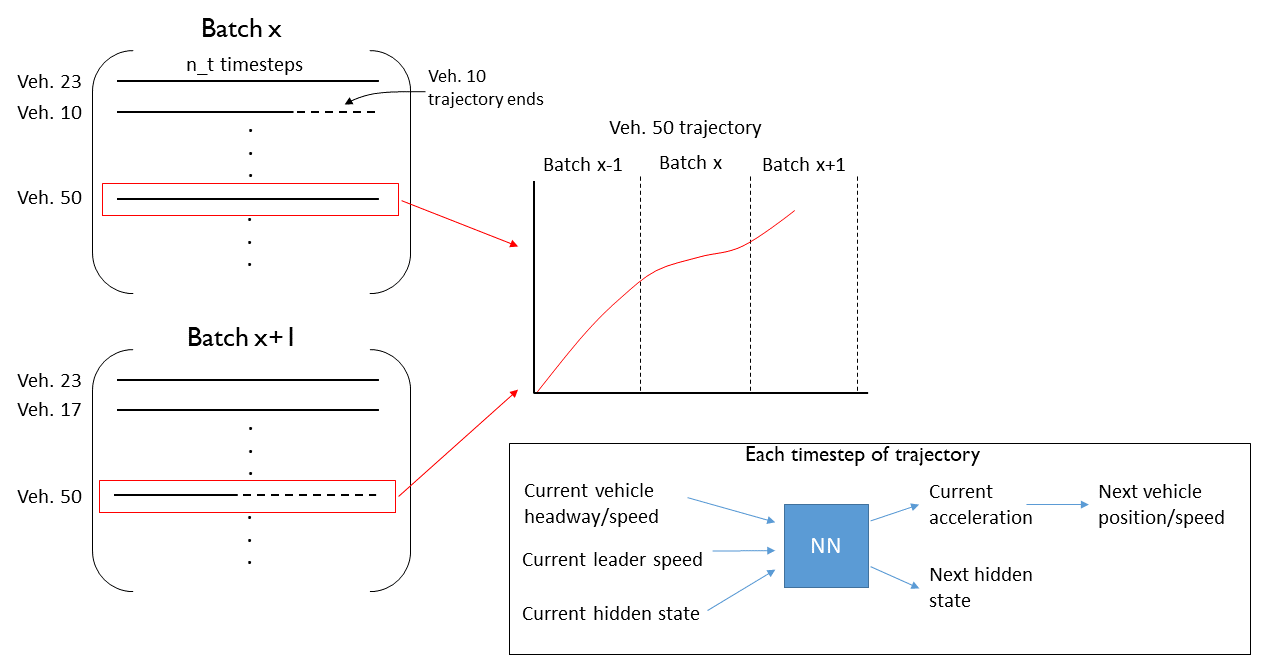} 
\caption{Illustration of batching and inputs/outputs of the LSTM model.}  \label{NNfig}
\end{figure}

To test if the relaxation model can be applied to nonparametric models, we tested a long short-term memory (LSTM) neural network based car following model. LSTM is a common type of recurrent neural network architecture which is designed to process time series data. Our methodology is primarily based off of \cite{zhou-rnn} and is also similar to other works such as \cite{huang-rnn, zhang-rnn}. \\
At every timestep, the current headway, vehicle's speed, and leader's speed are normalized and fed as inputs. The current hidden state of the LSTM, which is initialized as all zeros, is also fed as input. The neural network consists of one layer with 60 LSTM units connected to a second dense layer with 10 neurons. The LSTM produces a new hidden state, and the dense layer then connects to the output, which is a single float value representing the current acceleration of the vehicle. The outputted acceleration then updates the vehicle's position/speed, and the new current hidden state is recorded. \\
Note that this follows the approach used in \cite{zhou-rnn} where the same hidden state is constantly updated along the entire length of the vehicle trajectory, which typically lasts around 500-700 timesteps (50-70 seconds). In this way, the LSTM layer only has to process one timestep worth of data per prediction. In contrast, \cite{huang-rnn} generates a new hidden state for each timestep by feeding the past $M$ timesteps as an input (where $M$ is a hyperparameter taken as 50 timesteps). \\
The main difference in our neural network is how we handled the batching. In existing papers using deep learning for traffic modeling, they typically just state the batch size without elaborating further. This leads to confusion: are entire trajectories being simulated during training, do they just predict the next timestep given the past empirical trajectory? We found that the latter typically lead to bad testing results. If the former is used, do the batches consist of sequential timesteps for the same vehicle, or are multiple vehicles considered in each batch? We formed batches by selecting $n\_veh$ vehicles randomly. Each vehicle in the batch will have up to a maximum of $n\_t$ timesteps simulated, starting with the initial condition. 
In the next batch, the next $n\_t$ timesteps will be simulated for each of the $n\_veh$ vehicles, starting from where the previous batch's simulation ended. Once a vehicle's entire trajectory has been simulated, it is replaced with a new random vehicle in the next batch. This is explained visually in the Fig. \ref{NNfig}. Compared to having a fixed batch size, the final testing error was significantly improved by starting with a small value of $n\_t$ (50) before increasing it to around 500, so entire trajectories will be simulated in each batch. \\
Other minor details are as follows. We didn't clean or filter the data ourselves, and we used every vehicle possible in the reconstructed NGSim dataset. Vehicles were put into either the training (85\%) or testing (15\%) set, and the loss of the testing set is computed after training to ensure there is no overfitting. A dropout rate of .3 was applied to the LSTM layer, and l2 regularization with constant .02 was applied to the kernels of the dense and LSTM layers. The LSTM layer used a tanh activation and sigmoid recurrent activation, and the dense layer used ReLU activation. Adam with a learning rate of .0008 was used as the optimizer, using MSE in position as the loss function.
\subsubsection{Results of the LSTM model}
Applying relaxation to the LSTM model is the same as applying it to a parametric model, with one exception: when a lane change occurs, we pass the relaxation amounts as an extra input. Otherwise, the extra input is zero. We did this so there is a way for the neural net to `see' when a lane change occurs. The relaxation parameter was not learned during training, rather we treated it as a hyperparameter which was set as 12 manually.
\begin{table}[H]
\begin{tabular}{|l|l|l|l|l|l|l|}
\hline
                    & MSE ($m^2$) & MSE near LC & Realistic Acc.  & MSE many LC & MSE merges \\ \hline
LSTM relax   & 52.0/24.1/120   & 99.4/26.8/343  & 85\% & 88.7/34.5/232 & 37.4/15.8/71.9 \\ \hline
LSTM relax no LC    & 39.0/18.9/64.8  & -  & 100\% & - & -  \\ \hline
LSTM no relax   & 77.5/28.0/189   & 132/29.3/458  & 99\% & 144/48.8/344 & 42.6/13.3/82.3\\ \hline
LSTM no relax no LC   & 32.4/15.1/54.9  & -  & 100\% & - & - \\ \hline
\end{tabular} 
\caption{Calibration results of the unrelaxed/relaxed neural network (LSTM) model.}\label{table 2}
\end{table}
First, it should be stated that these MSE are much higher than those reported for the parametric models - this is not a surprise, as here we are learning a single set of parameters to predict the trajectory of any vehicle in the NGSim data. That is, the LSTM model encodes the behavior for the average driver, as opposed to having vehicle specific parameters. \\
Regarding the distributions of MSE, they all follow the similar patterns observed for the parametric models. The MSE is skewed right, with the error primarily dominated by vehicles which change lanes several times. We see that when training with relaxation, the prediction error on lane changing vehicles is significantly reduced (from 77.5 to 52. on average), but even with the relaxation added the error for lane changing vehicles is still significantly higher than for vehicles with no lane changing. For the LSTM with relax, the error on vehicles with no LC is slightly higher than the LSTM with no relaxation, however the LSTM with relax has a significantly lower error when considering the prediction over all vehicles (46.0 v 56.8). Thus, the relaxed LSTM model is more successful at learning the behavior of the average driver. LSTM with relax had a lower realistic acceleration metric, but upon visual inspection of the trajectories, we found this was caused by having accelerations just outside the [4, -6] bounds, as opposed to the trajectories being appearing qualitatively unrealistic.\\
Overall we found that the relaxation model is general enough to be easily applied to a neural network model, and that it had a significant improvement on the prediction of car following trajectories with lane changing.

\section{Macroscopic Experiments and Validation}

\begin{figure}[H] 
\centering 
\includegraphics[ width=.6\textwidth]{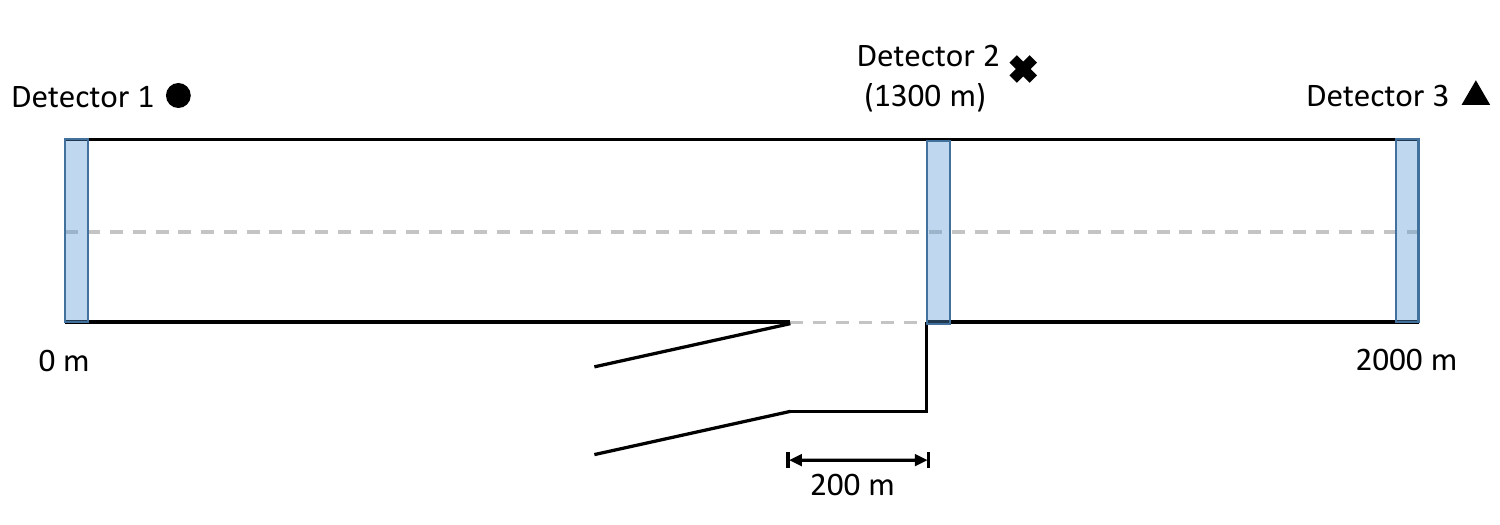} 
\caption{Configuration of road network used for macroscopic experiments.} \label{roadnetwork}
\end{figure}

\begin{figure}[H] 
\centering 
\includegraphics[ width=\textwidth]{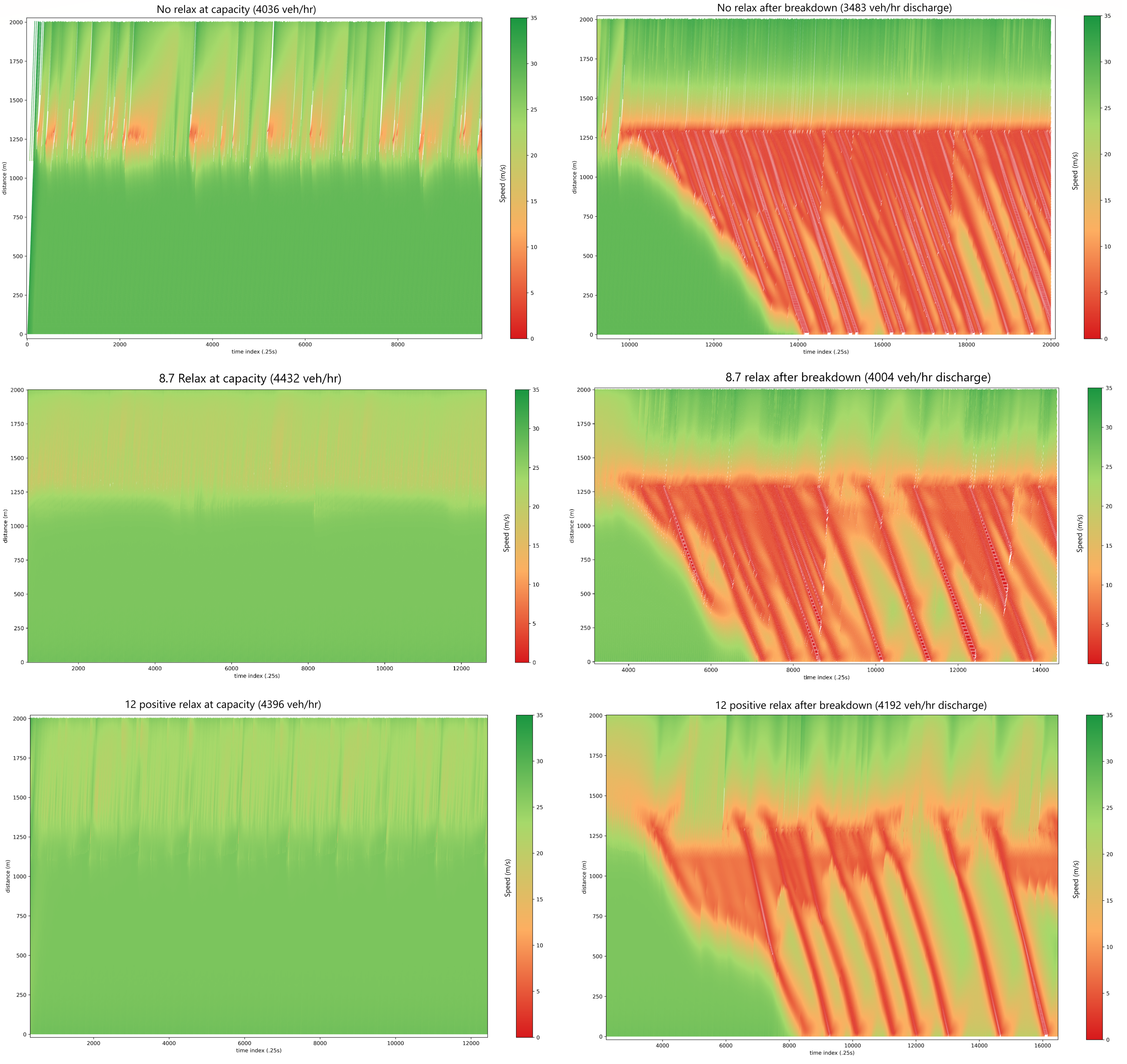} 
\caption{Space-time trajectories of the right lane (connected to the on-ramp). Trajectories are colored according to their speeds. 
} \label{macrofig1}
\end{figure}

To study the effects of relaxation on capacity drop, we considered the 2 lane highway with an on-ramp bottleneck shown in Fig. \ref{roadnetwork}. Simulations were done using our open source simulator havsim. The current havsim model is most similar to the open source simulator traffic-simulation-de \cite{TreiberKestingBook}, and it is explained fully in the appendix. Note that our model is completely deterministic, except for discretionary LC, and that we used identical vehicles in the simulations.
The reason for capacity drop at a merge bottleneck can be explained intuitively in the context of microsimulation. While traffic is uncongested and below capacity, vehicle trajectories immediately downstream of the bottleneck will be close to equilibrium (i.e. vehicles tend to have a fairly constant speed when leaving the bottleneck). When the capacity drop occurs, it is because merging vehicles cause disturbances on the main road--- this forms waves which propagate upstream on the main road, and the downstream trajectories will no longer be in equilibrium. It is well known that the equilibrium solution corresponds to a higher flow state than an oscillatory solution, and so because of the oscillations induced by merging vehicles, the discharge rate is reduced. Note that this intuitive explanation is in general agreement with work such as \cite{leclerqcapacitydrop} and \cite{duretcapacitydrop}, which have analyzed capacity drop in the Newell model with bounded acceleration. However, the key difference is that whereas the reduced flow in the Newell trajectories is due to empty space (voids) in the downstream trajectories, for a generic second order car following model the reduced flow is due to the oscillatory downstream trajectories. \\
In our experiments, we found that the relaxation time parameter had a strong effect on the oscillations created by merging vehicles. The Fig. \ref{macrofig1} shows the contrast between none, moderate, and strong relaxation. To produce those figures, we kept a constant inflow of 400 veh/hr on the on-ramp, and found the maximum constant inflow we could send on the main road before traffic breakdown would occur. The top left shows the maximum flow state for the model without relaxation. Even though the equilibrium solution corresponding to the maximum flow gives 4420 veh/hr, the model is only able to produce a maximum of 4036 veh/hr discharge. The relaxed model (bottom/middle left), are able to reach 4420 veh/hr, and its downstream trajectories show significantly less oscillation. The right figures show the unrelaxed/relaxed models after traffic breakdown. The unrelaxed model has more frequent and severe waves emerging from the merge, resulting in a significantly lower discharge rate. As the relaxation increases, oscillations tend to decrease, leading to higher discharge rates and lower values of capacity drop. 

One important aspect of capacity drop is that its magnitude depends on the amount of flow on the merging lane. This result has been found in \cite{leclerqcapacitydrop} for an analytic model as well as in \cite{duretmetering} and \cite{integrationdrop} for the microsimulators Aimsun and Integration respectively. Intuitively this is not too surprising: if there are more merging vehicles, there will tend to be more disturbances/oscillations, so the capacity and discharge rate both decrease (the capacity drop will increase). However we were not able to find any empirical studies which considered the amount of on-ramp flow as a key exogenous factor in capacity drop. \\
The effects of changing the relaxation time and inflow amounts are summarized in the two tables \ref{table 4} and \ref{table 5}. In those tables, period refers to the average time between shockwaves. We calculated the period by recording the arrival times of waves at the detector 1 (upstream). 
\begin{table}[H]
\centering
\begin{tabular}{|l|l|l|l|l|l|}
\hline
   Relax   (s)         & capacity & discharge (left/right) & drop \% (stdev)& period (stdev) (min.)\\ \hline
   0 & 4036 & 3483 (1851/1632) & 13.7 (2.9) & 1.73 (.82)  \\ \hline
    2 & 4432 & 3567 (1895/1672) & 19.5 (3.3) & 1.73 (.69)  \\ \hline
     4 & 4432 & 3710 (1924/1786) & 16.3 (3.1) & 1.89 (.96) \\ \hline
      7 & 4432 & 3859 (2001/1859) & 12.9 (3.1) & 2.88 (.86)  \\ \hline
      10 & 4432 & 4015 (2054/1961) & 9.4 (3.3) & 3.15 (1.39)  \\ \hline
        15 & 4396 & 3985 (2039/1947) & 9.3 (2.8) & 3.2 (1.43)  \\ \hline
        $10^*$ & 4432 & 4119 (2121/1998) & 7.1 (1.3) & 3.71 (1.02)  \\ \hline
\end{tabular}  
\caption{Results of changing relaxation time with constant on-ramp inflow of 400 veh/hr}\label{table 4}
\end{table}

\begin{figure}[H] 
\centering 
\includegraphics[ width=\textwidth]{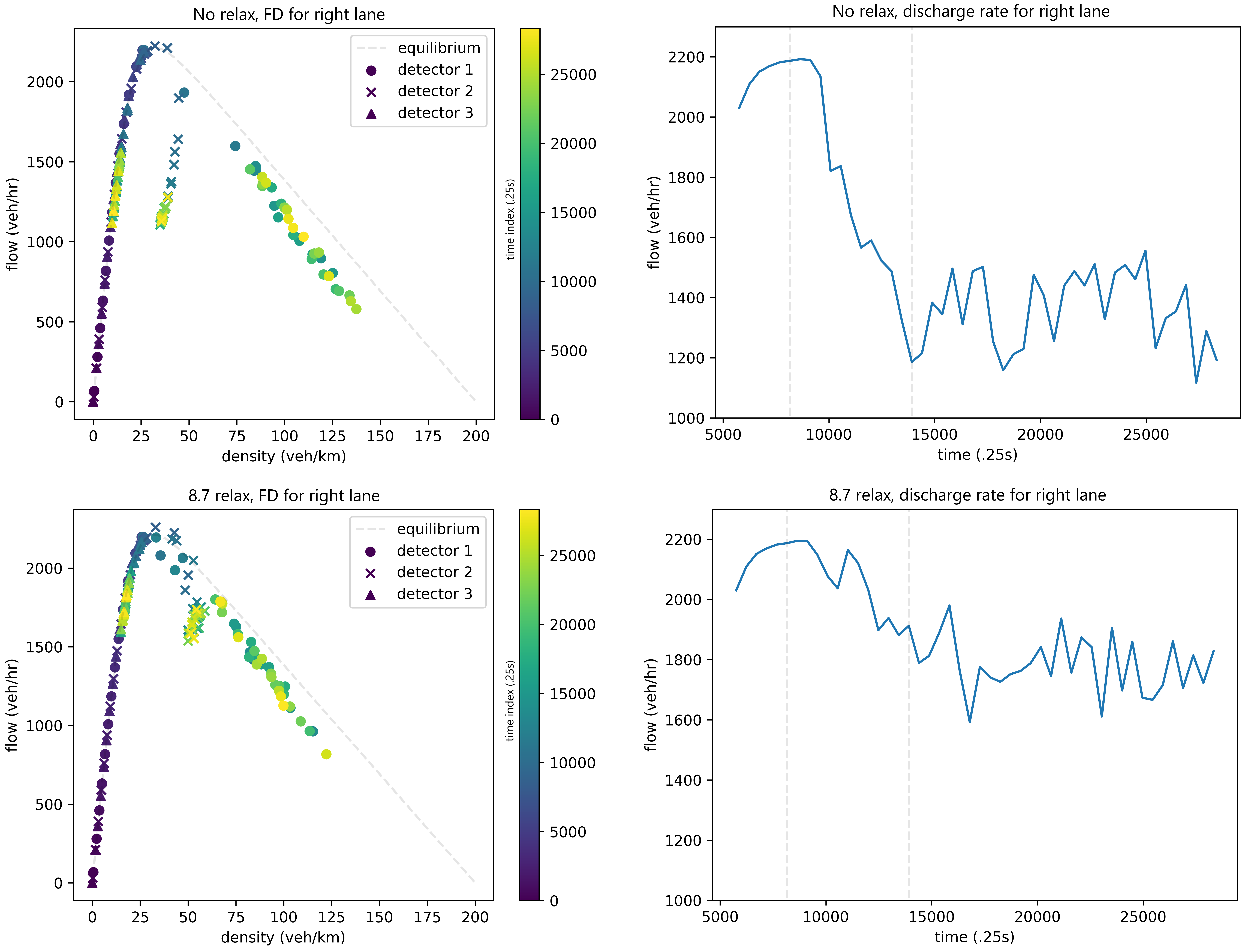} 
\caption{In left plots, the dashed line corresponds to the equilibrium solution of the car following model; the colors represent the time the measurement was taken. In right plots, the two dashed lines are at 34 and 58 minutes (during this period, the on-ramp inflow was gradually increased). }  \label{macrofig2}
\end{figure}

\begin{table}[H]
\centering
\begin{tabular}{|l|l|l|l|l|}
\hline
   Relax    (s)        & capacity & discharge (left/right) & drop \% (stdev)& period (stdev) (min.)\\ \hline
   0 & 3608 & 3110 (1720/1390) & 13.8 (4.5) & .88 (.40)  \\ \hline
    2 & 4040 & 3211 (1735/1476) & 20.5 (4.0) & 1.02 (.43)  \\ \hline
     4 & 4112 & 3305 (1723/1583) & 19.6 (4.3) & .98 (.40)  \\ \hline
      7 & 4148 & 3412 (1765/1647) & 17.7 (3.8) & 1.21 (.56) \\ \hline
      10 & 4256 & 3718 (1889/1829) & 12.6 (3.8) &  1.56 (.77)  \\ \hline
        15 & 4256 & 3676 (1881/1795) & 13.6 (4.1) & 1.62 (.55)  \\ \hline
        $10^*$ & 4184 & 3947 (2006/1941) & 5.7 (3.6) & 2.19 (1.28) \\ \hline
\end{tabular}  
\caption{Results of changing relaxation time with constant on-ramp inflow of 800 veh/hr}\label{table 5}
\end{table}
Note that there is some ambiguity in defining the capacity drop. As is pointed out in \cite{coifmanrelax} (and observed in our simulations), the discharge rate will increase above capacity shortly before breakdown. Further, since the capacity depends on the on-ramp inflow, the observed capacity drop will be different if the on-ramp inflow varies. The capacity and discharge rates were calculated as in the Fig. \ref{macrofig1}, where we keep the on-ramp inflow constant and vary the mainroad inflow until traffic breakdown; the capacity is then defined as the maximum inflow we could send before breakdown. To measure the discharge, we record the flow after breakdown at detector 3 in 2 minute intervals, averaged over $\approx$45 minutes. \\
We used our vanilla formulation which applies relaxation to both positive and negative $\gamma$. We also tried only applying positive relaxation, i.e. $\gamma >0$, as is done in other relaxation formulations, and those results are marked by an asterisk. The table \ref{table 6} shows the corresponding TTE and DT, which we estimated for a follower with an initial headway of 15 m and speed of 29 m/s, whose leader has a constant speed of 29 m/s (equilibrium headway is 55 m in this case). Those values were chosen to represent a merge in uncongested conditions; in a different merging situation, the TTE and DT would be different.
\begin{table}[H]
\centering
\begin{tabular}{|l|l|l|l|l|l|l|}
\hline
   Relax  (s) & 0 & 2 & 4 & 7 & 10 & 15  \\ \hline
 TTE/DT (s) & 24.3/1.8 & 25.5/3.5 & 26.7/5.4 & 28.5/8.2 & 30.3/10.9 & 33.4/15.4 \\ \hline
\end{tabular}  
\caption{Estimates of time to equilibrium (TTE) and deceleration time (DT) for IDM with varying relaxation time.}\label{table 6}
\end{table}
Overall this shows that increasing the relaxation time leads to higher discharge rates and longer periods between shockwaves. The capacity tends to remain fairly constant, meaning the capacity drop decreases as relaxation increases. Decreasing the negative relaxation (i.e. relaxation for larger gaps after LC) seems to have a similar effect as increasing the positive relaxation. Changing only the relaxation parameter, we produced capacity drop for a 2 lane highway in the range of $\approx$10-20\%, with shockwave periods between $\approx$1-5 minutes, which is in good agreement with empirical data \cite{capacitydropreview}.\\
Lastly, we show some examples of the FD and discharge rate in Fig. \ref{macrofig2}. In those experiments, we first generated the uncongested branch of the FD by keeping the on-ramp inflow at 0, and linearally increasing the main road inflow from 0 to 2196 veh/hr/lane over 24 minutes. The 2196 veh/hr/lane inflow is then kept constant for the rest of the experiment. Then from 34-58 minutes, the on-ramp inflow is increased linearally from 0 to 800 veh/hr, and kept constant for the rest of the 2 hour simulation. Points in the FD are aggregated over 2 minute time intervals, and measured over 100m distances using Edie's generalized definitions of flow/density; the discharge rate shows the flows of the downstream detector (3). We used a value of 8.7 for the relaxation time, which was the median value calibrated for IDM in the previous section. The fundamental diagrams show that 700m downstream of the merge, trajectories have returned to equilibrium. Immediately downstream of the merge, trajectories can be far from equilibrium, with flow values approximately equal to or slightly lower than the discharge rate, but with higher densities.
\section{Future outlook} \label{conclusion}
This work shows the importance of considering lane changing (LC) dynamics when formulating and and analyzing microsimulation models. LC dynamics are necessary to be able to understand and explain the trajectories of lane changing vehicles. More importantly, LC dynamics such as relaxation play a significant role in key traffic phenomena such as the formation of waves and capacity drop. We believe traffic flow researchers need to move towards a more holistic view of traffic microsimulation, which considers not only car following models, but also LC dynamics and mandatory/discretionary LC. To this end, we have proposed a relaxation model that can be applied to an arbitrary car following model, and made our complete microsimulation model available online at \url{https://github.com/ronan-keane/havsim}. That repository also includes the trained LSTM model and the code to generate the other results of the paper. \\
In the future, we hope that a general tactical/cooperation model can be developed, with the goal of obtaining a complete LC dynamics model which can be applied to an arbitrary car following rule. Since many traffic waves seem to originate from large perturbations caused by lane changing, as opposed to phantom traffic jams, a better understanding of LC dynamics would likely lead to better control strategies for improving traffic flow.

\section*{Appendix - havsim microsimulation model for highways}
Note that our model is almost entirely deterministic, as the only source of randomness is from discretionary LC. The car following, safety condition, and discretionary model follow traffic-simulation-de \cite{TreiberKestingBook}. This is combined with our relaxation and tactical/cooperation model, with modifications to the discretionary/mandatory LC model and upstream boundary conditions of traffic-simulation-de. 
\subsubsection*{Car following}
The IDM (Eq. \eqref{6}) is used with parameters $[c_1, c_2, c_3, c_4, c_5] = [35, \  1.3, \ 2, \ 1.1, \ 1.5]$
\subsubsection*{Lane changing model notation}
The ego vehicle $x_i$ is the vehicle for which the lane changing model is being evaluated. $x_i$ has the current leader, follower vehicles $x_{i-1}, x_{i+1}$ respectively. If $x_i$ were to change lanes, it would have the vehicles $x_{i-1}^*, x_{i+1}^*$ as the new leader, follower; we call $x_{i-1}^*, x_{i+1}^*$ the LC side leader, follower, respectively. LC side refers to either the left or right lane, depending on which side the ego vehicle is evaluating the change for. The current leader/follower relationships are refered to as the vehicle order, and if this order were to change due to lane changing, we call the new leader/follower relationships the new vehicle order.
Lastly, define $s(x_{i}, x_{i-1})$ as the space headway between vehicles $x_i$ and $x_{i-1}$, and define $h(x_i, x_{i-1})$ as the car following model evaluated for vehicle $x_i$ with $x_{i-1}$ as the leader.
\subsubsection*{Gap acceptance/safety condition}
The safety of the LC is evaluated by calling the car following model for the ego vehicle and LC side follower, under the new potential vehicle order. Thus the LC side follower safety is defined as $h(x_{i+1}^*, x_i)$ and ego vehicle safety is $h(x_i, x_{i-1}^*)$. For the potential change to be evaluated as safe, both safety conditions must be greater than the threshold value 
$$h(x_{i+1}^*, x_i) \text{ and } h(x_i, x_{i-1}^*) > d_1\frac{\dot x_i}{v_{\rm max}} + d_2(1 - \frac{\dot x_i}{v_{\rm max}})$$
where $v_{\rm max}$ is the maximum possible speed of $x_i$ (for the IDM, the $v_{\rm max} = c_1$), and $d_1, d_2$ are parameters taken as $-8, -20$. 
\subsubsection*{Discretionary LC and incentive}
For a discretionary change to be accepted, the safety condition and incentive condition must both be met. The incentive condition follows the MOBIL, defined as 
$$ h(x_i, x_{i-1}^*) - h(x_{i}, x_{i-1}) + d_4\left(h(x_{i+1}, x_{i-1}) - h(x_{i+1}, x_i) + h(x_{i+1}^*, x_i) - h(x_{i+1}^*, x_{i-1}^*)\right) + d_5 > d_3$$
where $d_3$ is the incentive threshold, $d_4$ is the politeness, and $d_5$ is the bias for the left side ($d_6$ is the bias for right side). Those parameters are taken as $.6,\  .1,\  0,\  .2$ respectively. \\
While in a discretionary state, there is a $d_7 = .1$ probability to check the incentive and safety conditions in any given timestep. If a discretionary change is completed, the discretionary model will not be checked for the next $d_9 = 20$ timesteps.  
\subsubsection*{Mandatory LC}
When vehicles on the on-ramp are able to merge onto the main road, they enter a mandatory LC state. In this state they will check the safety condition every timestep, and change as soon as the safety conditions are met. 
\subsubsection*{Tactical/Cooperation model}
The tactical and cooperation model is applied a) when the vehicle is in a mandatory state and one or both of the safety conditions are not met or b) the vehicle is in a discretionary state and the incentive is met but not the safety conditions. In case b), the discretionary state becomes ``activated'' so that the LC model will be continuous evaluated for the next $d_8=20$ timesteps (as opposed to the normal case, when there is only a $d_7$ probability to check the model). \\
Our tactical and cooperation model is very simple and based on either adding acceleration or deceleration to the baseline car following acceleration. A vehicle can either have $a2=2$ acceleration added or $a_3 =  -2$ deceleration added. Note that this has the effect of shifting the equilibrium of the baseline model. \\
In the cooperative model, the ego vehicle seeks cooperation from either the LC side follower or the LC side follower's follower. The headway between the potentially cooperating vehicle and the ego vehicle must be greater than the jam spacing of the cooperative vehicle. In the discretionary state, the cooperative vehicle will only choose to accept the cooperation with probability $a_1 = .2$. In the mandatory state, vehicles will always be willing to cooperate with the ego vehicle. If the LC side follower's safety is violated, the cooperating vehicle will have deceleration added to its baseline car following acceleration, so that it will give a greater amount of space. \\
In the tactical model, the ego vehicle adjusts its acceleration so that it will position itself better. If the LC side follower's safety is violated, the ego vehicle always has acceleration added; otherwise, only the ego vehicle's safety is violated, so deceleration is added to the ego vehicle.
\subsubsection*{Relaxation}
Follows the formulation of section 2.0 and 2.2.
\subsubsection*{Boundary Conditions}
The downstream boundary condition is applied whenever a vehicle lacks a leader. We used a free boundary, which is equivalent to taking the limit of the car following model when the leader position goes to infinity and leader speed goes to $v_{\rm max}$. For the IDM, this yields
$$\ddot{x}_i = c_4\left( 1 - \left( \dfrac{\dot x_i(t)}{c_1} \right)^4\right)$$
For the upstream boundaries, each lane with inflow has an ``inflow buffer'' which is incremented according to the instantaneous flow rate at any given timestep. When the inflow buffer $>=1$, we attempt to add a vehicle to the simulation at the start of the corresponding lane. The vehicle will potentially be added with an initial speed $v_0$
$$v_0 = \max\{\dot x_{i-1}, v_{\rm eql}^i(s(x_i, x_{i-1})) \} $$
provided that the headway satisfies
$$s(x_i, x_{i-1}) >= b^* s_{\rm eql}^i(v_0)$$
$$b^* = \begin{cases} 
 b_1 & v_0 > b_2 \\ 
  1 & \text{o.w.}
  \end{cases}$$
 Where we use the notation $v_{\rm eql}^i$ and $s_{\rm eql}^i$ to refer to the equilibrium solution for vehicle $i$. 
 $b_1$ and $b_2$ are parameters controlling the boundary condition. $b1 < 1$ allows vehicles to be aggresively added to the simulation. Under congested conditions, vehicles tend to be added with an initial speed of $b2$. We took $b_1=.8$ and $b_2=18.85$ (18.85 is the speed giving maximal flow for the chosen parameters of IDM).

\section*{Acknowledgement}
Thank you to the anonymous reviewers for their helpful comments. Thank you to Nathan Bala and Yidan Wang for working with us on havsim. \\

\noindent The contents of this work reflect the views of the authors, who are responsible for the facts and the accuracy of the information presented herein. This document is disseminated in the interest of information exchange. The work is funded, partially or entirely, by a grant from the U.S. Department of Transportation’s University Transportation Centers Program. However, the U.S. Government assumes no liability for the contents or use thereof.

\bibliographystyle{unsrt}
\bibliography{mysources1}

\end{document}